# Optimized motor imagery paradigm based on imagining Chinese characters writing movement

zhaoyang Qiu, Brendan Z. Allison, Jing Jin, *Member, IEEE*, Yu Zhang, *Member, IEEE*, Xingyu Wang, Wei Li, Andrzej Cichocki, *Fellow, IEEE*

*Abstract*—*Background:* Motor imagery (MI) is a mental representation of motor behavior that has been widely used as a control method for a brain–computer interface (BCI), allowing communication for the physically impaired. The performance of MI based BCI mainly depends on the subject's ability to self-modulate EEG signals. Proper training can help naive subjects learn to modulate brain activity proficiently. However, training subjects typically involves abstract motor tasks and is time-consuming. *Methods:* To improve the performance of naive subjects during motor imagery, a novel paradigm was presented that would guide naive subjects to modulate brain activity effectively. In this new paradigm, pictures of the left or right hand were used as cues for subjects to finish the motor imagery task. Fourteen healthy subjects (11 male, aged 22–25 years, mean 23.6±1.16) participated in this study. The task was to imagine writing a Chinese character. Specifically, subjects could imagine hand movements following the sequence of writing strokes in the Chinese character. This paradigm was meant to find an effective and familiar action for most Chinese people, to provide them with a specific, extensively practiced task and help them modulate brain activity. *Results:* Results showed that the writing task paradigm yielded significantly better performance than the traditional arrow paradigm (p<0.001). Questionnaire replies indicated that most subjects thought the new paradigm was easier and more comfortable. *Conclusions:* The proposed new motor imagery paradigm could guide subjects to help them modulate brain activity effectively. Results showed that there were significant improvements using new paradigm, both in classification accuracy and usability.

*Index Terms*—motor imagery, brain-computer interface (BCI), CSP, paradigm.

This work was supported in part by the Grant National Natural Science Foundation of China, under Grant Nos. 61573142, 61203127, 91420302 and 61305028. This work was also supported by the 13 Fundamental Research Funds for the Central Universities (WG1414005, WH1314023, and WH1516018) and Shanghai Chenguang Program under Grant 14CG31.

Z. Qiu, Yu Zhang, *X. Wang and *J. Jin are with the Key Laboratory of Advanced Control and Optimization for Chemical Processes, Ministry of Education, East China University of Science and Technology, Shanghai, P.R. China (*correspondence e-mail: xywang@ecust.edu.cn, jinjingat@gmail.com).
B.Z. Allison is with Guger Technologies OG, 8020 Graz, Austria.
W. Li is with Shenyang Institute of Automation, Chinese Academy of Sciences, Shenyang 110016, China, with the School of Electrical Engineering and Automation, Tianjin University, Tianjin 300072, China, and also with the Department of Computer & Electrical Engineering and Computer Science, California State University, Bakersfield, California 93311, USA (corresponding author, e-mail: wli@csub.edu).
A. Cichocki is with Laboratory for Advanced Brain Signal Processing, Brain Science Institute, RIKEN, Wako-shi, 351-0198, Japan and Systems Research Institute of Polish Academy of Science, Warsaw, Poland.

## I. INTRODUCTION

Brain-computer interface systems (BCIs) can translate brain activities into commands that could be used to control external devices [1-3]. BCIs provide a new communication channel for people, which is different from communication channels that rely on the conventional neuromuscular pathways of peripheral nerves and muscles [4]. Many BCIs are based on the EEG recorded noninvasively via electrodes placed on the scalp. While various different neural activities can be used as features in electroencephalogram (EEG) based BCIs, most BCI systems rely on motor imagery [5, 6], event-related potential [7-10] or visual evoked potentials [11-14].

It has been well established that imagination of limb movement could result in event-related desynchronization (ERD) [5] and event-related synchronization (ERS) [6]. Specifically, movement imagery (MI) affects the mu (8-12Hz) and beta waves (13-30Hz) in the EEG. Movement imagery produces significant ERD over the contralateral central area during imagination of right and left hand movement [15]. The basic principle of an MI based BCI entails translating ERD phenomena into control commands [16, 17]. However, the differences are often not sufficient to provide reliable control signals in practical applications [18-20]. It has been reported that about 20% of MI BCI users do not attain sufficient accuracy to control an application [18]. This phenomenon was initially called "BCI illiteracy", although other terms such as "BCI inefficiency" [21] have also been used.

Many researchers have endeavored to develop algorithms to improve performance [22, 23]. Channel selection methods have been widely used in MI based BCIs to enhance performance by removing task irrelevant and redundant channels [24-26]. Some advanced machine learning algorithms were set up to improve the performance of BCI [27, 28]. These classifiers could yield high classification accuracy. In addition, various feature extraction methods have been proposed to detect ERD/ERS patterns, which could discriminate different mental states effectively [29-31]. However, no matter how outstanding the algorithms are, BCIs cannot obtain good performance if subjects cannot modulate brain activity as required by the BCI [16].

To help users modulate brain activity proficiently, many feedback training methods have been proposed to improve the performance of MI-based BCIs. Feedback can help subjects master effective strategies to modulate brain activity by providing them the information about EEG changes. In [32], Neuper et al. compared realistic feedback and abstract feedback for the same task. Results showed that there was not much difference in training performance. Blankertz et al. proposed the Berlin Brain-Computer Interface (BBCI), in which feedback was provided by a discrete form [33]. The BBCI could exhibit effective performance quickly in untrained subjects. Yue et al. (2012) presented an EEG-based real-time control paradigm with visual feedback, in which the external device was a simulated inverted pendulum [34]. With suitable training procedures, subjects would learn to control unstable devices through BCI. [35] investigated the effect of static and dynamic visual representations of target movements during the BCI neurofeedback training. Other work introduced an auditory feedback method, and the results showed that the auditory feedback was a suitable substitute for the visual equivalent [36]. However, it has been reported that feedback can have both inhibitory and facilitative effects on EEG control [37]. Each normal feedback training session should begin with a calibration run without feedback. The calibration data are employed to construct the classifier for the next feedback runs. Hence, feedback performance largely depends on the initial model, and the calibration data may not be sufficiently discriminable. If subjects were unfamiliar with the system and failed at MI tasks at the initial state, the feedback training session could frustrate subjects [38].

At the beginning of training process, naive subjects may try various motor imagery strategies. If the MI based BCI system can provide subjects a proper cue that instructs them to do motor imagery the right way, it will save training time and reduce frustration. Therefore, it is necessary to set up a suitable paradigm for subjects, which will help them to perform the MI tasks effectively.

In most paradigms, only visual or auditory cues were used to instruct subjects to perform the corresponding classes of MI tasks (such as left hand movement, right hand movement or foot movement) [34, 39]. But these cues could not instruct subjects how to perform the imagination. The type of movement imagery can have a major effect on BCI performance, for example, [40] found that subjects who relied on visual imagination of hand movement performed worse than subjects who focused on the kinesthetic aspect of the same task. [41] found that subjects could attain better performance by imagining familiar actions.

The strategy of motor imagery in [41] was effective for a small group of people who had formal musical training and had played and practiced piano regularly. Since different subjects were familiar with different actions, the purpose of this paper was to find a familiar action that can be accepted by most Chinese people. Therefore, we developed a novel paradigm to guide subjects to imagine Chinese characters writing movements with their right or left hand.

In this paradigm, a picture of a hand was used as a cue for subjects to focus on one of their hands. A Chinese character shown on the screen was used to guide subjects regarding what kind of motor imagery task they should perform. Subjects imagined moving the chosen hand to follow the strokes of the shown Chinese character. In other words, the subjects imagined that they were writing on the blackboard using the right or left hand. This paradigm would give subjects a specific MI task instruction: moving from left to right, moving from top to bottom, drawing an oblique line or a point, etc. The task of writing a Chinese character can make people focus more attention on motor imagery. We compared this task to a traditional approach in which an arrow cue stimulus was used.

## II. METHOD

### A. Subjects

Fourteen healthy subjects (11 male, aged 22-25 years, mean 23.6±1.16) participated in the motor imagery experiment. All subjects signed a written consent form prior to this experiment and were paid for their participation. The local ethics committee approved the consent form and experimental procedure before any subjects participated. All subjects were right handed with no clinical history of neurological disorders, according to self-reports. All subjects' native language was Mandarin Chinese. None of them had experience with any MI based BCI before this study.

### B. Experimental paradigms

After being prepared for EEG recording, the subjects were seated in a comfortable chair about 80cm from a standard 24 inch LED monitor (60 Hz refresh rate, 1920×1080 screen resolution) in a shielded room. The cues were presented in the middle of the screen. During data acquisition, subjects were asked to relax and avoid unnecessary movement.

Two different types of paradigms were utilized and compared in this study: (1) the traditional arrow paradigm, (2) the writing task paradigm.

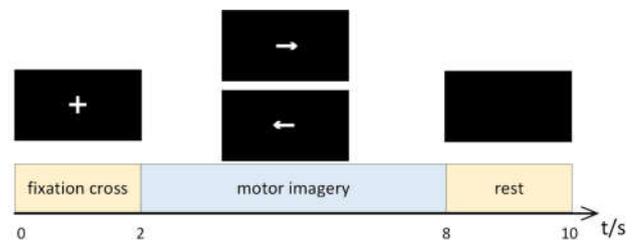

Fig. 1. Timing of a trial of the traditional arrow paradigm. Each trial consisted of task and rest periods. Subjects started to execute MI tasks while the arrow appeared on the screen. The time segment between seconds 3 and 7 was used for feature extraction.

Throughout both paradigms, motor imagery was performed without feedback. The subjects were told to focus on motor imagery during the experiment. For each subject, two classes of motor imagery were performed: left hand and right hand. These tasks were pseudo randomly distributed throughout the run. Each experimental paradigm included two runs, and each run is comprised of 50 single trials. A total of 100 trials of EEG measurements were performed (half for each class of MI).

(1) The traditional arrow paradigm: This paradigm used an arrow pointing to the left or right, which has been widely used [42, 43]. As illustrated in Fig. 1, at the beginning of a trial (t=0s), a fixation cross appeared on the black screen. After two seconds (t=2s), an arrow cue was shown that pointed either the left or the right (corresponding to one of the two classes of MI tasks: left hand or right hand). The subjects had 6s to imagine the movement of either their left or right hand, as indicated by the cue arrow. In this paradigm, each subject was asked to imagine a hand movement which they felt was easiest for them, and so the exact types of motor imagery were not consistent across subjects. A short break followed, during which the screen was black.

(2) The writing task paradigm: Fig. 2 presented the writing task paradigm. Unlike the traditional arrow paradigm, a static photograph of a Chinese character and a forearm was displayed on the black screen. Before the motor imagery (t=0s), a fixation cross was shown on the black screen instructing subjects to get ready. After two seconds (t=2s), the fixation cross was replaced by a picture of the forearm and Chinese character. The picture of the forearm was used as a cue to tell subjects which hand movement they should imagine. The Chinese character was to instruct the subjects what kind of the writing movement they should imagine. Each subject had 6s to perform the MI task. During this period, the subjects were asked to imagine hand movements that followed the strokes of the Chinese character on the screen. It was as though the subject was writing the character on the blackboard with the right or left hand. A short break (2s) followed, during which the screen was black. Each Chinese character used in this paper contains 5 strokes.

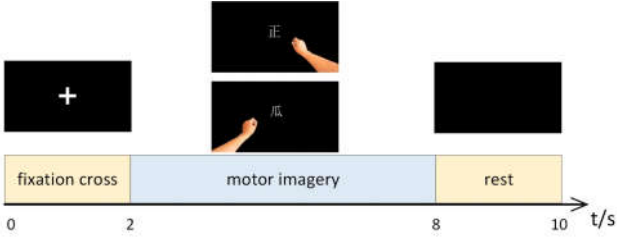

Fig. 2. Timing of a trial of the writing task paradigm. Each trial consisted of task and rest periods. Subjects started to execute MI tasks while the character and a forearm appeared on the screen. The time segment between seconds 3 and 7 was used for feature extraction.

### C. EEG acquisition

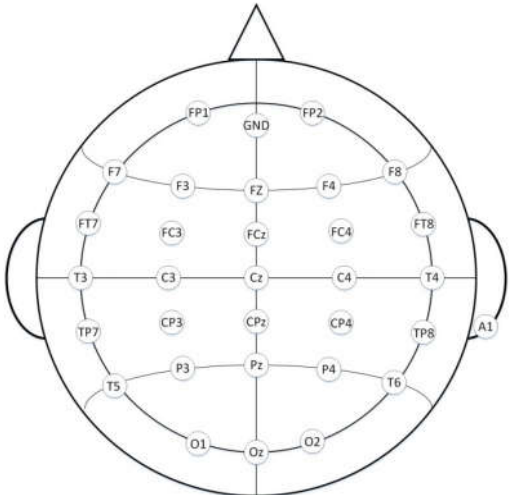

Fig. 3. The electrode distribution used in this study. All channels were used for signal recording and analysis.

EEG signals were sampled at 250 Hz through the Neuro Scan amplifier (high-pass and low-pass filters: 0.1 Hz and 50 Hz, with the mains power supply (50Hz) filtered out) using a 32-channel cap following the 10-20 international system (Fig. 3). All channels were used for signal recording and analysis, which were referenced to electrode A1 located over the right mastoid with a forehead ground (GND). All impedances were below 5 k ohms. The sequence of the two paradigms in this experiment was counterbalanced.

### D. Feature extraction procedure

For motor imagery feature extraction, the 30 channels shown in Fig. 3 were used without further channel selection. The EEG data were band-pass filtered using a fifth order Butterworth band pass filter from 8 to 30 Hz, since this frequency band included the range of frequencies that were mainly involved in performing motor imagery.

The CSP algorithm is an efficient feature extraction algorithm, which has been widely used in MI based BCI systems [44-47]. CSP is based on the simultaneous diagonalization of two covariance matrices. It finds a spatial filter to maximize variance for one class and minimize variance for another class at the same time to improve classification.

### E. Classification scheme

A support vector machine (SVM) is a machine learning method proposed in the 1990s [48]. It is mainly proposed for two class pattern recognition problems. For a given data set $A \in R^d$ from two classes that can be divided by a hyperplane linearly, the hyperplane is expressed as: $WA + b = 0$. $W \in R^d$ is weight vector and $b$ is the intercept (scalar). The problem is transformed into finding the optimal hyperplane as follows:

$$\min \varnothing(W, \varepsilon) = \frac{1}{2} \|W\|^2 + c \sum_{i=1}^{n} \varepsilon_i, c \geq 0 \qquad (10)$$

$$s.t.\ y_i(W^T A^{(i)} + b) \geq 1 - \varepsilon_i, \varepsilon_i \geq 0, i = (1, \cdots n)$$

The optimization problem is a convex quadratic programming problem [49]. Where $A^{(i)}$ is a feature vector of a training sample, $y_i$ is the category with labels {-1, 1} which $A^{(i)}$ belongs to. $W$ is the hyperplane coefficients vector. The parameter $\varepsilon_i$ is called slack variables and $c$ is regularization parameter. c is used to control the trade-off between the model complexity and empirical risk [50]. 10*10-fold cross validations accuracy of the data were used for each subject.

### F. Subjective report

After completing each session, each subject was asked a question about the paradigm: Was this paradigm hard? The question could be answered on a 1-5 Likert scale indicating strong disagreement, moderate disagreement, neutrality, moderate agreement, or strong agreement.

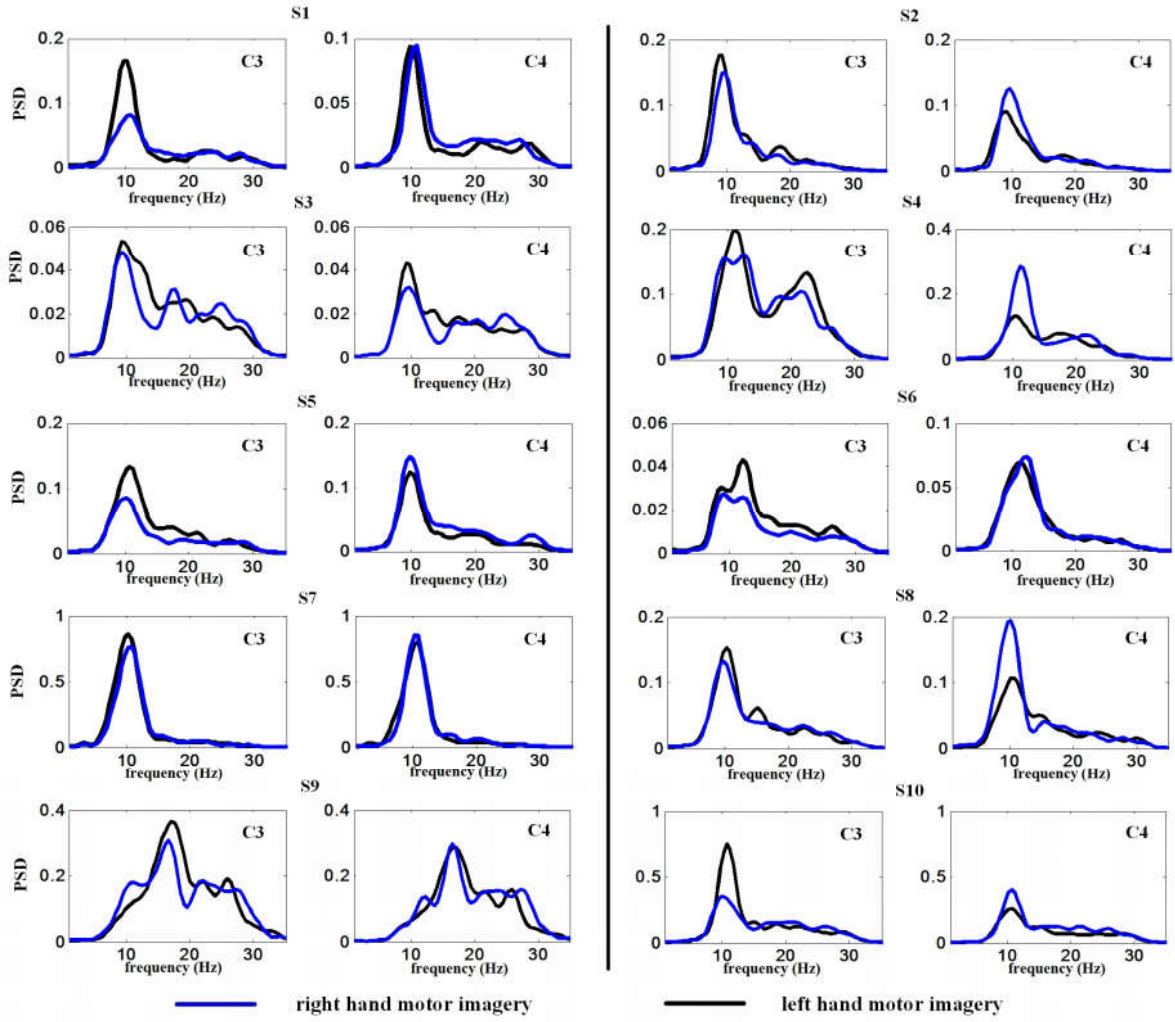

Fig.4. This figure displays the averaged spectra of 100 trials recorded at C3 and C4 for each subject in the writing task paradigm (blue: right hand, black: left hand). Lines in the map have been smoothed. S1-S10 refers to subjects 1-10.

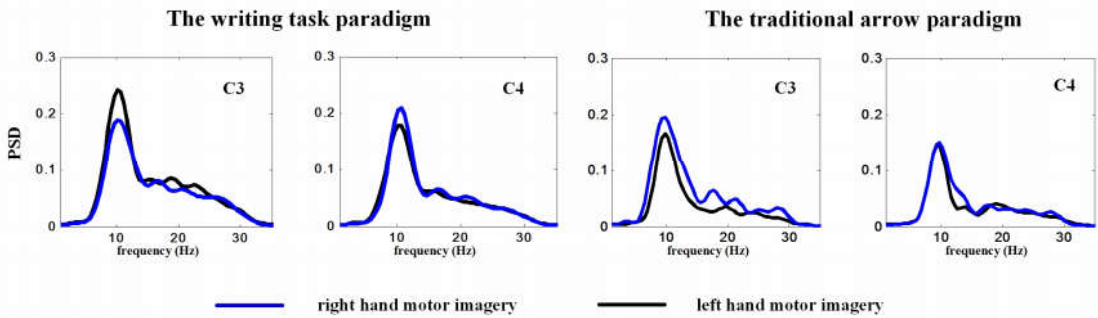

Fig.5. This figure displays the averaged spectra of two different paradigms at C3 and C4 (blue: right hand, black: left hand). For each paradigm, the maps are averaged across all ten subjects.

## III. RESULTS

Different articles have presented different thresholds for BCI "efficiency" [51]. Since we used subjects who were new to motor imagery BCIs, we used a low threshold. Four subjects attained accuracies less than 60% in both paradigms. These subjects were not selected for further analysis. Ultimately, results from ten subjects are shown below.

### A. EEG results

To illustrate the differences between two classes of motor imagery in the writing task paradigm, power spectral densities (PSDs) were calculated from the average data, plotted in Fig.4. Each map was based on 100 averaged trials (50 left hand motor imagery trials and 50 right hand motor imagery trials) after being filtered by 8-30 Hz band pass filter.

Fig. 5 presents the PSDs averaged across all ten subjects from two different paradigms at C3 and C4. For the writing task paradigm, the energy from right hand motor imagery was lower than left hand motor imagery at C3, and the energy from left hand motor imagery was lower than right hand motor imagery at C4. In the beta band, signals from right hand motor imagery were higher than left hand motor imagery at both channels. For the traditional arrow paradigm, signals from right hand motor imagery were higher than left hand motor imagery at C3, while there was no significant difference of two classes of MI tasks at C4.

In CSP method, W is the projection matrix, and $W^{-1}$ is the inverse matrix of W. The columns of $W^{-1}$ are the time invariant vectors of EEG source distribution vectors called common spatial patterns [52]. The first and last common spatial patterns are shown in Fig. 6. The first pattern was obtained by maximizing the variance of the right hand motor imagery, which was associated with the ERD phenomenon over the area of the left sensorimotor area of the cortex. Accordingly, the ERD phenomenon over the right motor area was associated with the last pattern, corresponding to the left hand motor imagery.

Results in Fig. 6 show that, for the writing task paradigm, spatial discriminability of right and left hand motor imagery was apparent. Maps also indicated that areas around channel C3 and C4 were associated with the left and right hand motor imageries, which was consistent with the neurophysiology phenomenon reported in [53-54]. Results of the traditional arrow paradigm did not show clear pattern from the contra lateral hemisphere for left and right motor imagery. ERS phenomena were also observed during left hand motor imagery in the writing task paradigm.

### B. Classification Performance Comparison

For each of the two paradigms, a 10×10-fold cross-validation was implemented to evaluate the classification performance. The classification accuracies of all ten subjects from both paradigms are shown in Fig. 7. Results showed that the classification accuracies of the new writing task paradigm were better than the traditional arrow paradigm for all subjects. The average classification accuracy of writing task paradigm was 79.8%, which was 14.7% higher than traditional arrow paradigm (65.1%). Among all ten subjects, S5 achieved the greatest improvement by using the writing task paradigm (increased by 26%).

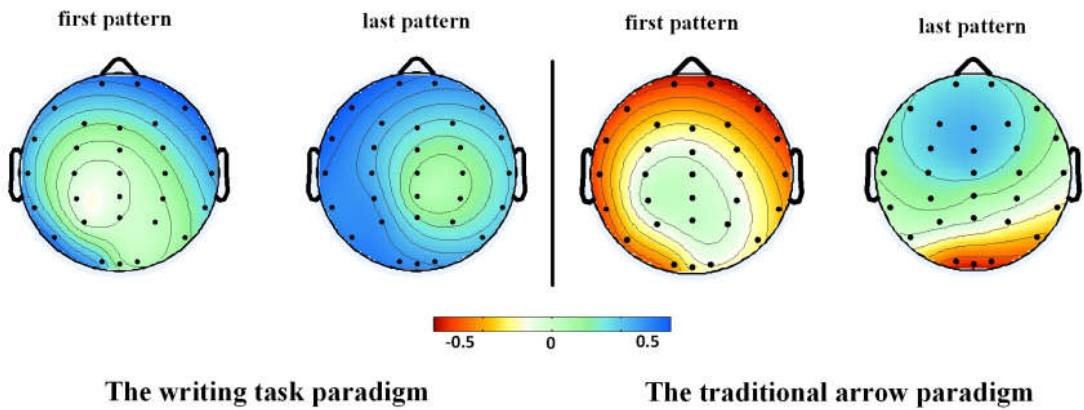

Fig. 6. The topographic maps of the first and last spatial patterns extracted by the CSP method. The patterns were extracted from averages of 100 trials of the two different paradigms.

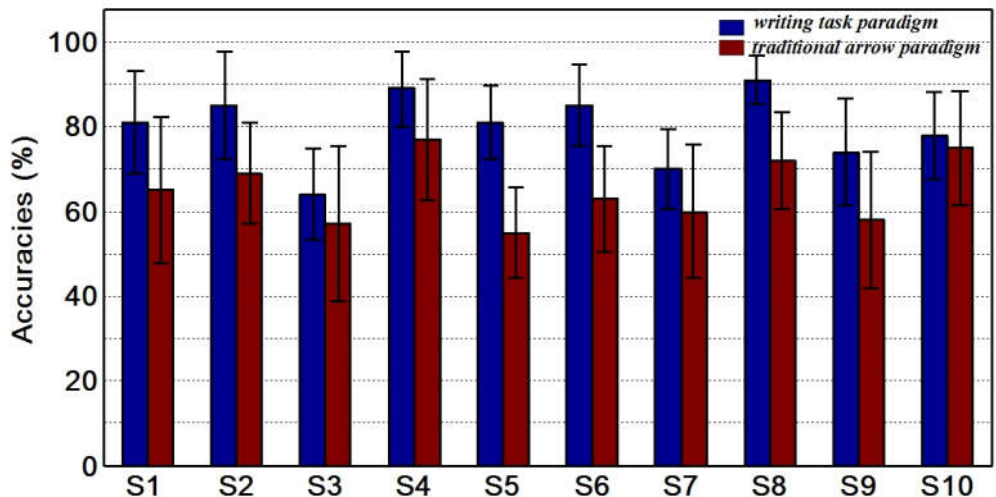

Fig. 7. Classification accuracies of all ten subjects from the writing task paradigm and traditional arrow paradigm.

TABLE I
TEN SUBJECTS' RESPONSES TO QUESTIONS ABOUT THE TWO PARADIGMS

|  | S1 | S2 | S3 | S4 | S5 | S6 | S7 | S8 | S9 | S10 | Average |
|---|---|---|---|---|---|---|---|---|---|---|---|
| The normal arrow paradigm | 2 | 4 | 5 | 2 | 4 | 3 | 3 | 4 | 5 | 4 | 3.6±1.1 |
| The writing task paradigm | 2 | 3 | 4 | 1 | 1 | 3 | 2 | 2 | 4 | 2 | 2.4±1.1 |

A paired samples T-Test was used to show the classification accuracy differences between the writing task paradigm and traditional arrow paradigm (p<0.001). It showed the classification accuracy was increased significantly by using our new paradigm compared to the traditional paradigm.

*C. Subjective report*

Ten subjects' responses to the question about task difficulty are shown in Table 1. The replies are based on a 1-5 Likert scale to indicate strong disagreement, moderate disagreement, neutrality, moderate agreement, or strong agreement. In the comparison of difficulty, significant differences were found between the two paradigms (p=0.0026).

IV. DISCUSSION

The primary goal of this study was to investigate whether the new paradigm presented in this paper could help subjects in a motor imagery (MI) BCI. Two paradigms were compared in our experiment. The results showed that the writing task paradigm yielded significantly better performance than the traditional arrow paradigm.

Fig.4 displays the averaged spectra of the two motor imagery tasks in the writing task paradigm. Several points stand out: (1) For channel C3, the power during right hand motor imagery was lower than left hand motor imagery in the band around 8-12 Hz. (2) For channel C4, the power during left hand motor imagery was lower than right hand motor imagery in the band around 8-12 Hz. Activity in the alpha-band (8-12 Hz) is called the Mu rhythm, which is recorded at rest and attenuated by voluntary movements. Research initially suggested that the Mu rhythm is an ''idling'' rhythm in the motor cortex [55, 56]. However, Mu rhythm activation has been also reported during the execution of different tasks. As reported in [43], the Mu-rhythm is task sensitive, and can change based on motor imagery in the motor cortex.

The traditional arrow paradigm instructs subjects to perform a corresponding MI tasks, and subjects imagined a movement they prefer. Subject S1, S2, S4, S5, S8 and S9 reported that they imagined grasping an object using the left or right hand in this paradigm. Subjects S3 and S10 imagined lifting a dumbbell. Subject S6 imagined shaking hands with other people. Subject S7 imagined boxing. Hence, the user-selected imagery was not consistent across the ten subjects in the traditional arrow paradigm. Many subjects reported that they thought that imagining the same movement every time would make them bored and tired, and could cause them to lose focus. Performing different tasks could reduce boredom. However, some subjects thought that it would be difficult to imagine different movements in different trials. In the writing task paradigm, different characters were displayed in different trials. This task was also very simple and highly practiced: imagining writing. Subjects' responses to the questionnaires indicated that they thought that the writing task paradigm could help them focus their attention and reduce fatigue. Table 1 shows that the subject S5 considered the new paradigm much easier and more comfortable. Concordantly, S5 exhibited the greatest difference between the two paradigms (26%), shown in Fig. 7. Table 1 indicates that S4 thought that both paradigms are easy, and he performed well in both paradigms (89% in the writing task paradigm and 77% in the traditional arrow paradigm).

Although this study involving 10 subjects yielded good performance using a novel motor imagery paradigm in a BCI, further work is needed. After the experiment, some subjects said that imagining writing with the left hand made them feel awkward (all subjects were right handed). Some other subjects said that they imagined writing the strokes in a different sequence when they wrote with the left hand. They prefer to imagine the characters as abstract symbols and copy them. Although this did not appear to affect their task performance, it will be interesting to find the differences in brain patterns for these different subject states.

Future work could also explore the comparison between user-selected imagery and experimenter-defined imagery. While the approach used by Wolpaw and colleagues encourages subjects to explore different mental strategies for control, the approach from Pfurtscheller and colleagues adopted by other groups generally does not. Future work could further study the relative merits of these two approaches to motor imagery for different users.

V. CONCLUSIONS

In this study, we introduced a new motor imagery based BCI paradigm using Chinese characters as visual cues, which we hypothesized would help subjects perform two MI tasks: writing the character with the right or left hand. Experimental results based on ten healthy subjects demonstrated that this new paradigm yielded significantly better classification accuracies than the conventional paradigm using arrows as visual cues (p<0.001). Furthermore, most subjects thought that the new paradigm was easier, based on their questionnaire responses. Future work will focus on simplifying the characters to shorten the time required to complete the imagined task, and experiments with online feedback.